\documentstyle[12pt]{article}
\evensidemargin\oddsidemargin
\begin{document}
\count0 = 1
\begin{titlepage}
\vspace {20mm}
\begin{center}
\ QUANTUM SPACE-TIME AND REFERENCE FRAMES \\
IN  GRAVITY AND FLAT SPACE-TIME\\
\smallskip
\vspace{6mm}
\bf{S.N. MAYBUROV}
\vspace{6mm}

\small{Lebedev Institute of Physics}\\
\small{Leninsky pr. 53, Moscow Russia, 117924}\\

\vspace{15mm}
%\small{\bf Abstract}
\end {center}
\vspace{3mm}
\begin {abstract}
%\small{
The quantum space-time model which accounts material 
  Reference Frames (RF) quantum effects  considered for flat space-time 
  and ADM canonical gravity. As was shown by Aharonov
 for RF - free material object its c.m. nonrelativistic motion
in vacuum described by Schrodinger wave packet evolution
which modify space coordinate operator of test particle in this RF
and changes its Heisenberg  uncertainty relations.
In the relativistic case we show that
 Lorentz transformations between two RFs   include
the  quantum corrections for RFs momentum uncertainty
 and in general can be formulated as
 the  quantum space-time transformations.
 As the result for moving RF  its Lorentz time boost acquires
quantum fluctuations  which calculated solving 
relativistic Heisenberg equations for the quantum clocks models.
It permits to calculate RF proper time for the arbitrary RF quantum motion
including quantum gravity metrics fluctuations.
  Space-time structure of canonical Quantum Gravity 
and its observables evolution for RF proper time discussed
 in  this quantum space-time transformations framework.
\end {abstract}
\small {Submitted to Int. J. Theor. Phys. }\\
\vspace {32mm}

\small {Talk given at 'Quantum Field Theory under External conditions'\\
 conference , Leipzig, September 1998\\
to appear in proceedings.\\
\\
\\
\\
\\}
\vspace{28mm}
\small {  * E-mail  Mayburov@sgi.lpi.msk.su}
\end{titlepage}
\begin{sloppypar}
\section{Introduction}
The possible changes of space-time properties at small (Plank) scale
now  extensively discussed $\cite{Aha,Dop}$. Due to the absence of
 any experimental 
information it seems instructive to look for some directions 
 exploring attentively the standard
Quantum Physics space-time structure.    
Some years ago Aharonov  and Kaufherr have shown that in nonrelativistic
Quantum Mechanics (QM) the correct definition of physical reference 
frame (RF) must differ from commonly accepted one, which in fact was
transferred copiously from Classical Physics $\cite{Aha}$. The main reason  
 is that to perform exact quantum description
 one should account the quantum properties 
not only of studied object, but also RF, despite the possible practical
 smallness.   
 The most simple  of this RF properties is the existence of Schroedinger
 wave packet of free macroscopic object with which RF is usually
associated $\cite {Schiff}$. Then it  introduces
additional uncertainty into the measurement of object space coordinate
 in this RF.
Furthermore  this effect  results in the states transformations
between two such RFs which includes quantum corrections to the
  standard Galilean group transfromations $\cite{Aha}$.
Algebraic and group theorettical structure of this transformation was
studied in $\cite{Tol}$.
In their work Aharonov and Kaufherr formulated Quantum Equivalence 
Principle (QEP) in nonrelativistic QM  - all the laws of Physics are invariant
 under transformations between both classic and this finite mass RFs
which called quantum RFs. 
  The importance of RF quantum properties account was shown already in 
Quantum Gravity and Cosmology studies  $\cite {Rov,Unr,Dew}$ and will
be considered here in connection with the time problem in quantum
gravity.

 In this paper the consistent
 relativistic covariant  theory of quantum RFs formulated,
 our preliminary results were published \cite{May2}.
In this theory no  new {\it ad hoc} hypothesis introduced;
 all  calculations  are performed in the standard QM formalism.
 It will be shown that the transformation of the  particle state 
 between two quantum RFs obeys to relativistic invariance
principles, but  differs from standard Poincare Group
transformations, due to quantum relativistic correction
for RF motion. Solving the evolution equation
for quantum clocks models the proper time in moving quantum RF 
calculated and the related effects of RF momentum quantum fluctuations 
revealed. This clocks model applied for the analysis of
 the space-time structure of   
canonical quantum gravity $\cite {Rov}$.

 In chap.2 canonical
formalism for quantum RFs described. In chap.3 we study
quantum clocks models and obtain relativistic proper time
for quantum RF. In chap.4 the relativistic evolution equations
and unitary transformations for quantum RFs described.
 We'll consider also RF quantum motion in gravitation field
where gravitational 'red shift' results in additional
clocks time fluctuations.
 
\section{Quantum Coordinates Transformations }

For the beginning we'll consider  Quantum Measurements problems related 
to Quantum RFs model.
 In QM framework the system defined as RF should be able to   measure 
the observables of studied quantum states and so
 include the measuring device - detector D.
 As the realistic  example  we can regard the photoemulsion plate or
the diamond crystal which can measure the particle position 
 and simultaneously record it.
  Despite the multiple proposals up to now the established
theory of  collapse  doesn't exist
  $\cite{May,Desp}$.
  Yet our problem premises doesn't connected
 directly with any state vector collapse mechanism and
 and it's enough to detailize standard QM collapse postulate of
von Neumann measurement theory $\cite {Desp}$.
 We consider  RF which consists of 
 finite number of atoms (usually rigidly connected)  and have the finite
 mass.
 It's well known that the solution of Schroedinger equation for
  any free quantum system  can be factorized as :
\begin {equation}
  \Psi(t)=
\sum c_l\Phi^c_l(\vec{R}_c,t)*\phi_l(u_{k},t) \label {A1}
\end {equation}
where  center of mass coordinate $\vec{R}_c=\sum m_i*\vec{r}_i/M$,
  $c_l$ are the partial amplitudes. $u_k$ describes the internal degrees of
freedom, which for potential forces are reduced to 
 $\vec{r}_{i,j}=\vec{r}_i-\vec{r}_j$
 $\cite{Schiff}$. Here $\Phi^c_l$ describes the c.m. motion of the system.
 It means that the evolution of the
 system  is separated into the external
evolution  of  pointlike particle M and the internal evolution
 defined by $\phi_l(u_{k},t)$.
 So the internal evolution is
 independent of whether the  system is localized
 in some 'absolute' reference frame (ARF)
 or not. Quantum Field Theory  evidences that the
 factorization of c.m.
 and relative motion holds true even for nonpotential forces and 
 variable $N$ in the secondarily quantized systems $\cite{Schw}$.
 Moreover this factorization expected to be correct for nonrelativistic
 systems 
 where binding energy is much less then its mass $M$, which is
 characteristic for the real detectors and clocks.
 Consequently it's reasonable
to assume that this factorization  fulfilled also
for the detector states despite we don't know
their exact structure. For our problem it's enough to assume 
  that   eq.(\ref{A1}) holds for RF state
  only in the time interval $T$
 from RF preparation moment $t_0$ until the act of measurement starts
, i.e. the measured particle $n$ wave packet $\psi_n$ impacts with D.  
  If this factorization holds the space coordinate measured in this RF 
depends not only on $\psi_n$ but also on $\Phi_l^c$ which permit
in principle to study quantum RF effects.  In this case
 the possible factorization violation at later time when
the particle state collapse occured is unimportant for us.
  We regard 
  in our model that all measurements  are performed on
the  quantum pairs ensemble of particles $G^2$ and
  $F^1$. It means that each event is resulted 
 from the interaction between the 'fresh' RF and particle
 ,prepared both in the specified quantum
 states, alike the particle alone in the standard experiment.
% To simplify our calculations  we'll  take normally that in initial
% $F^1$ state all $c_l=0$ except $c_1$.

To illustrate the meaning of Quantum RF
 consider gedankenexperiment  where in ARF
the wave packet of RF $F^1$  
  described by  $\psi_1=\eta_1(x)\xi_1(y)\zeta_1(z)$ at  time moment $t_0$. 
 The test particle $n$
  with mass $m_n$ belongs to   narrow beam which average velocity
is  orthogonal to $x$ axe
 and  its wave function at $t_0$ is $\psi_n=\eta_n(x)\xi_n(y)\zeta_n(z)$.
 Before they start
to interact this system wave function is the product of $F^1$ and $n$
packets.
We want to find $n$ wave function for the observer in
 $F^1$ rest frame. In general it can be done by means of the canonical 
transformations described  below, 
but if $n$ beam state is localized so that  
$\psi_n$ can be approximated
 by  delta-function $\delta(x-x_b)\delta(y-y_b)\delta(z-z_b)$
then $ n$ wave function in $F^1$ can be easily calculated
$\psi'_n(\vec{r}'_n)=\eta_1(x'_n-x_b)\xi_1(y'_n-y_b)\zeta_1(z'_n-z_b)$.
 It shows that if for example $F^1$ wave packet along $x$ axe
have average width $\sigma_x$ then from the 'point of view'
of observer in $F^1$ each object localized in ARF acquires wave packet of the
same width $\sigma_x$ in $F^1$ and any measurement  in $F^1$ and ARF
 will  confirm this conclusion.
 
  The generalized Jacoby canonical  formalism will be applied in our
model alternatively to  Quantum Potentials  used in $\cite{Aha}$.
 Consider the system $S_N$ of $N$ objects $W^k$   which
 include $N_f$ frames $F^i$ which
 have also some internal degrees of freedom and
 $N_g=N-N_f$ pointlike 'particles' $G^i$.
At this stage we can regard both of them as equivalent objects in the relation
to their c.m. motion.
 We'll assume for the beginning  that particles and RFs canonical 
 operators $\vec{r}_i,\vec{p}_i$ are defined
 in absolute (classical) ARF - $F^0$ having very large mass $m_0$,
 but later this assumption can be abandoned.
We'll start with Jacoby canonical coordinates  $\vec{u}_j^l$  associated
with  $F^l$ rest frame, which for $l=1$ equal    :
\begin{eqnarray}
 \vec{u}^1_{i}=\frac{\sum\limits^N_{j=i+1}m_j\vec{r}_{j}}{M_{i+1}}
-\vec{r}^l_{i} ;\quad 1\le i<N ;
 \quad \vec{u}_N=\vec{u}_s=\vec{R}_{c} \quad \label{B1}
\end{eqnarray}
where  $M_i=\sum\limits^{N}_{j=i}m_j$. $\vec{u}^l_i$ can be  obtained
and is the linear combination of $\vec{u}^1_i$.
 Conjugated to $\vec{u}^l_i$ canonical momentums are  :
\begin {equation} 
\vec{\pi}^1_i=
\mu_i(\frac{\vec{p}^s_{i+1}}{M_{i+1}}-\frac{\vec{p}_{i}}{m_{i}}) ,\quad
\vec{\pi}_N=\vec{p}_s=\vec{p}^s_1 \label{B2}
\end {equation}
where $\vec{p}^s_{i}=\sum\limits^{N}_{j=i}\vec{p}_j$
,and reduced mass
 $ \mu^{-1}_i=M_{i+1}^{-1}+m_{i}^{-1} $ .
The relative coordinates  $\vec{r}_j-\vec{r}_1$ can be represented
 as the linear sum of
 $\vec{u}^1_i$. They don't
constitute canonical set due to the quantum motion of $F^1$ $\cite{Aha}$.
 The  Hamiltonian of $S_N$ motion  in ARF is expressed also via
momentums $\vec{\pi}^1_i$   :
\begin {equation} 
\hat{H}=\sum\limits_{i=1}^{N}\frac{\vec{p}_i^2}{2m_i}=
\frac{\vec{p}_s^2}{2M}+
\sum\limits_{j=1}^{N-1}\frac{(\vec{\pi}^{1}_j)^2}{2\mu_j}
=\hat{H}_s+\hat{H}_c   \label {B5}
\end {equation}
In $F^1$ rest frame the true observables are  $\vec{\pi}^1_i ,
\vec{u}^1_i$ and it's impossible to measure
$S_N$ observables  $\vec{p}_s$ and $\vec{R}_c$. The true 
Hamiltonian of $S_N$ in $F^1$ should depend on  the true observables
only , so we can regard $\hat{H}_c$ as the real candidate for its role.
It results into modified Schroedinger equation
 which depends not only of particles masses ,but on observer mass $m_1$
also. 

Now we'll regard here the alternatve form of this  formalism which use
 Jacoby frame condition (JFC) and
is more convenient for the relativistic problem. 
For the described system $S_N$  Langrangian in ARF
 $L=\sum\frac{m_i\dot{\vec{r}}_i^2}{2}$ 
  gives $H$ of ($\ref{B5}$) after Legandre transform.
If one wish to include ARF motion in this formalism the simplest
way is to define formally $L'=L+\frac{m_0\dot{\vec{r}}_0}{2}$.
 It gives $N+1$ canonical momentums :
 $\vec{p}_j=\frac{\partial L'}{\partial\dot{\vec{r}}_j}$.
The new Langrangian $L'$ is formally symmetric relative to 
 the frame choice and it gives the Hamiltonian $H'=H+\frac{\vec{p}_0^2}{2m_0}$ 
for $H$ of ($\ref{B5}$). Due to it to anchor this momentums 
 and $H'$ to  $F^i$ rest frame in which they 
 acquire some values
 one must broke $L'$ symmetry  introducing the frame condition (FC)
or kinematical (holonomial) constraint $\cite{Git}$. For ARF rest frame 
we choose FC $\vec{p}^{\,2n}_0 \approx 0$, where from the formal reasons
$n=2$. It means that $\dot{\vec{r}}_0=0$ - RF is at rest relative to
itself which seems quite natural, yet it
 differs from FC used in $\cite{Aha}$.
All Classical and QM  results are reproduced in this scheme if ARF mass
is taken infinite.
%We can use $\vec{\pi}^1_i,\vec{u}^1_i$of($\ref{B2}$) even for finite $m_0$. 
$S_N$ quantization in $F^1$ performed with Hamitonian $\hat{H'}$ 
and FC regarded as the operator which obeys to
 Dirack rules for the first order constraints $\cite {Dir,Git}$.
 Galilean-like passive transformations from ARF to  $F^1$
and back can be found introducing FC also for
  $F^1$  $\vec{p}^{\,2n}_{11}\approx 0$, where
 $\vec{p}_{1i}$ are the canonical momentums in $F^1$.
 $S_{N+1}$ unitary transformation from ARF to $F^1$ 
is convenient to write via the  $F^{0,1}$ total momentum
 $\vec{p}_f=\vec{p}_0+\vec{p}_1$
and $F^0,F^1$ relative momentum $\vec{\pi}_f$
 conserving other momentums $\vec{p}_i$.  
Their conjugated coordinates $\vec{r}_f,\vec{u}_f$ have the standard
form of ($\ref {B1}$). 
In this notations the transformation from $F^0$ to $F^1$ is equal to  :
\begin {equation}
      U_{1,0}=P_f e^{i a_f\vec{p}_f\vec{r}_f}e^{-i\vec{p}_f\vec{b}_s}
\prod_{i=2}^{N}e^{-im_i\vec{r}_i\vec{\beta}}   \label{BBB}
\end {equation}
where
 $a_f=\ln\frac{{m}_0}{m_1}, \vec{b}_s=\frac{M}{m_0}\langle\vec{R}_c\rangle$.
  $P_f$ is $\vec{r}_f$ reflection (parity) operator. 
 $\vec{\beta}=\frac{\vec{p}_f}{m_1}$ is the operator  corresponding
to the velocity parameter in Galilean transformation.
Under this transformation $\vec{p}_f$ transformed to 
 $\vec{p}_{1f}=\vec{p}_{10}+\vec{p}_{11}$
and  $\vec{\pi}_{1f}= \vec{\pi}_{f}$.
Alike the transformation from $\vec{p}_j$ to $\vec{\pi}^1_i$ obtained
operator $U_{1,0}$ includes the dilatation transformation $\cite{Bar}$. 

 For $N=2$  one obtains $F^1$ momentums and coordinates :
\begin{eqnarray}
\vec{p}_{10}=(1-\frac{m_0}{m_1})\vec{p}_0-\frac{m_0}{m_1}\vec{p}_1 \quad ;\quad
\vec{r}_{10}=-\frac{m_1\vec{r}_1+m_2\vec{r}_2}{m_0}+\vec{b}_s  \nonumber \\
\vec{p}_{11}=-\vec{p}_0 \quad ;\quad
\vec{r}_{11}=-\vec{r}_0+(1-\frac{m_1}{m_0})\vec{r}_1-\frac{m_2}{m_0}\vec{r}_2
+\vec{b}_s \label {BA} \\
\vec{p}_{12}=-\frac{m_2}{m_1}(\vec{p}_0+\vec{p}_1)+\vec{p}_2 \quad ; \quad
\vec{r}_{12}=\vec{r}_2  \nonumber
\end{eqnarray}
Results for $N>2$ can be easily deduced from this formulaes. 
It's easy to see that ARF FC transformed into  $F^1$ FC. All $\vec{\pi}^1_i$
are conserved and space shift on $\vec{b}_s$ conserves all the
distances $ \vec{r}_i-\vec{r}_j$.  
In the limit where heavy $F^1$ moves  nearly classically $U_{01}$ becomes 
the Galilean momentum transformation with the velocity
 $\langle\vec{\beta}\rangle$.
$S_{N+1}$ Hamitonian in $F^1$ also can be rewritten via
new relative momentums $\vec{\pi}_{1j}$ which can be easily
derived following ($\ref {B2}$)  :
\begin{equation}
\hat{H}^1=\sum^N_{i=0}\frac{\vec{p}^2_{1i}}{2m_i}=H^1_s+H^1_c
=\frac{\vec{p}_{1s}^2}{2M_{N+1}}+
\sum_{j=2}^{N+1}\frac{\vec{\pi}_{1j}^2}{2\mu_{1j}}
 \label {BYY}
\end{equation}
  The   term $\hat{H}^1_s$ describes
 $S_N$ c.m. motion relative to $F^1$ which 
  doesn't influence on the evolution of $S_{N+1}$ true observables
  $\vec{\pi}_{1i} , \vec{u}_{1i}$ or  $\vec{r}_i-\vec{r}_j$. 
$\vec{r}_{1i},\vec{p}_{1i}$ aren't $S_{N+1}$ observables
for $F^1$ observer, yet $\vec{p}_{1i}$ expectation values
can be found from $\vec{\pi}^1_i$ measurements.%doesn't conserve state vector norm, multiplying it by some functions of
%$m_i$, but we neglect its renormalization, because it can
%be related completely to unobservable $\vec{r}_{1i}$ state vectors
%components.

Now we have quantum system $S_{N+1}$ which include ARF and in ARF
 rest  frame we can ascribe to it 
without any contradictions with QM  the state vector which for $N=2$
is equal :
$\psi_s(\vec{p}_0,\vec{p}_1,\vec{p}_2)=\varphi(\vec{p}_1,\vec{p}_2)
|\vec{p}_0=0\rangle |\vec{p}_1\rangle |\vec{p}_2\rangle$.
After $U_{1,0}$ transformation it acquires the similar form in $F^1$
rest frame with $|\vec{p}_{11}=0\rangle$.
As the result of this transform we obtain the new canonical coordinates
referred to finite mass  $F^1$ rest frame. They permit to factorize 
internal $S_N$ motion and ARF motion and dropping ARF term in $H^1$
of ($\ref{BYY}$) we obtain $S_N$ Hamitonian.
Remind that active transformation shifts $G^2$ state $\psi_2$ on the
 distance $\vec{a}$
and velocity $\vec{\beta}$ relative to RF. Passive $G^2$ transformation
 means the transition from one RF to another, but for quantum RF with 
state $\psi_s$ it can't be described by any state shift on
 $\vec{a},\vec{\beta}$ and have more complicated form.
$U_{1,0}$ is such passive transformation and active $G^2$ transformation
is the standard Galilean one even in $F^1$  $\cite {Schw}$.

 In general the quantum transformations in 2 or 3 dimensions 
should also take into account the possible
 rotation of quantum RF axes relative to ARF, which introduce 
additional angular uncertainty into objects coordinates.  Thus after
 performing coordinate transformation $\hat{U}_{A,1}$ from ARF
to $F^1$ c.m.  we must rotate all the 
objects (including ARF) around it on the uncertain  polar and
azimuthal angles $,\phi_1,\theta_1$ which are $F^1$ internal
degrees of freedom. We can imagine $F^1$ axes as some
solid rods which orientation this angles describe.
 As the result the complete transformation is:
 $\hat{U}^T_{A,1}=\hat{U}^R_{A,1}\hat{U}_{A,1}$.
Such rotation transformation operator commutes with $\hat{H}_c$ and due to it
can't change the evolution of the transformed states $\cite{Aha,May2}$.

\section  {Quantum  Clocks Models}

To construct the relativistic covariant formalism of quantum RFs
it's necessary first to define the time in such RFs. 
In nonrelativistic mechanics time $t$ is universal and 
 is independent of observer, while in relativistic
case  each observer in principle has its own proper time $\tau$.
We don't know yet the nature of  time , but phenomenologically 
 it can be associated  with the clock hands motion or some other relative
motion of the system parts $\cite {Hol}$.
In Special Relativity the time in moving frame  $F^1$ can be defined
by external observer at rest  measuring the state of $F^1$ comoving clocks. 
We'll consider the same procedure in relativistic QM i.e. 
some clock observable being measured at some time from the rest frame  
 gives the estimate of  proper time of moving quantum RF $F^1$.

For some clocks models $F^1$ internal evolution which define $F^1$ clocks
 motion and consequently its proper time $\tau_1$ can be factorized
 from $F^1$ c.m. motion.
 Its quantum c.m. motion described  by the
 relativistic Schrodinger equation for massive boson.
 This is Klein-Gordon
square root (KGR) equation in which only positive root
will be regarded for initial positive energy state $\cite{Schw}$.
 Solving Dirack constraints 
it was shown recently that this first order equation is 
completely equivalent to
free Field secondary quantization $\cite {Git2}$.
  
For our relativistic model we should regard more strictly the features of
reference frames and clocks, taking into account the internal motion.
Consider the evolution of some system $F^1$ where the  internal interactions
described by the Hamiltonian $\hat{H}_c$ 
are nonrelativistic , which as was discussed  in chap.1 is a reasonable
 approximation for the measuring devices or clocks.
We'll use  the parameter $\alpha_I=\frac{\bar{H}_c}{m_1}$ 
,where $m_1$ is $F^1$ constituents total rest mass. In $F^1$ c.m. 
 $\alpha_I=m_1^{-1}\langle\varphi_c|\hat{H}_c|\varphi_c\rangle$
 where $\varphi_c$ is $F^1$ internal state of ($\ref{A1}$). It describes
the relative strength of the internal $F^1$ interactions and for the
realistic clocks is of the order $10^{-10}$.
 In addition we'll assume that all RF 
constituents spins and orbital momentums are  compensated 
so that its total orbital momentum is zero, like in $\alpha$-particle
ground state.
 In this case the system $F^1$ c.m. motion can be reduced to the
motion of the spinless boson with the mass  $m_1$ and
in the next order  the mass operator $m_t=m_1+H_c$ will be used.
We'll start the proper time study with the simple models
of quantum RFs with clocks, yet we expect its main results to be true
also for the more sophisticated models.

To introduce our main idea let's regard the dynamics of the
moving clocks in Special relativity $\cite {Dew}$. We'll suppose that the
proper  (clocks) time is defined by the coordinate $\theta$
describing  some internal system motion independent of its
c.m. motion. For the simplicity assume that 
 Hamiltonian of clocks  $H_c$ results in the trajectory
 $\theta(t)=\omega t+\theta_0$ of the clocks canonical observable $\theta$,
which  renormalized into the time observable $\tau=\frac{\theta}{\omega}$.
This is the property which is expected from ideal clocks and
the simplest example of such system is the motion of free
 particle relative to observer $\tau=\frac{x}{v}$ $\cite {Hol}$. 
 For this and some other clocks models described below the
Hamitonian of clocks
with mass $m_1$ which c.m. moves with momentum $\vec{p}_1$ relative 
to ARF   :
$$
      H_T=(m_t^2+\vec{p}_1^2)^\frac{1}{2}
$$
where $m_t=m_1+H_c$.
If $\theta ,\vec{p}_1$ commutes, solving Hamilton equations
in ARF time $\tau_0$ one obtains $\theta(\tau_0)=B_1\omega \tau_0+\theta_0'$,
where $B_1=\frac{m_t}{H_T}$ coincides with Lorentz boost value.
So as expected , if $\theta$ is measured by the observer at rest
he finds the proper time $\tau_1=B_1 \tau_0$ of moving frame.
Yet we'll show  that  the quantum fluctuations
of RF motion results in the principally new additional effects.
  
One of the most simple and illustrative
 quantum clocks  models is the quantum rotator  proposed by Peres
\cite{Per}.  The rotator  Hamiltonian 
 $\hat{H}_c=-2\pi \omega i \frac{\partial}{\partial\theta}$
, where $\theta$ is the rotator's  polar angle.
 Preparing the special  initial state
 $\varphi_c(\theta)=|v^0_J\rangle$ at $t=0$,
 where $J$ is its maximum orbital momentum
one obtains the close resemblance of
  the classical clocks hand motion. 
The clocks state $\varphi_c(\theta-2\pi\omega t)$
 for large $J$ has the sharp peak at
$\bar{\theta}=2\pi\omega t$ with the uncertainty
 $\Delta_{\theta}=\pm \frac{\pi}{N}$ and can be visualized as the constant 
hand motion on the clocks circle.

 Our main clocks model - $C_x$ exploits   
 the nonrelativistic particle  motion relative to observer with  Hamitonian
 $H_c=\frac{\vec{p}^{\,2}}{2m}$
 \cite {Hol}. Let's consider the particle 3-dimensional motion, but choose
 as its initial state at $t=0$
   the Gaussian packet factorized in $x$ direction which
momentum state vector is :
\begin {equation}
\phi_c(\vec{p})=A \phi(p_y,p_z)
e^{\frac{\sigma_x^2}{2}(\bar{p}_x -p_x)^2}  \quad \label {CD}
\end {equation}
for which $\bar{p}_x\ne 0$. $\sigma_x$ is the initial wave packet spatial
 spread. Then the simplest Hermitian
observable which gives the time estimate 
is  $\hat{\tau}=\frac{m x}{\bar{p_x}}$ -
the particle's position on the arbitrary $x$ axe.
It describes the nonshifted  measurement with $\bar{\tau}=t$ and
 the finite dispersion $D_0(t)$ for $0<t<\infty \cite {Hol}$.
In fact in $C_x$ model $\hat{\tau}$ is the clocks hand position operator
or the pseudotime operator, and not a time operator in a strict sense
 $\cite{Hol,Per}$.
So from all sides $C_x$ can be regarded as the realistic clocks model
 in which measuring $\hat{\tau}$ one obtains the correct $t$ estimate
 with some statistical error having quantum origin. 
$C_x$ wave function $\varphi_c(x,t)$ evolution can be factorized
as the packet centre of gravity motion with the constant velocity 
$\frac{p_x}{m}$
and the packet smearing around it. For the given initial
state  there is unambiguous correspondence 
between the state vector
$|\varphi_c(x,t)\rangle$ and time $t$, so the quantum clocks synchronization
at $t=0$ means the preparation of the state $\varphi_c(x,0)$.
 From the corresponding Heisenberg equation one can find 
 Heisenberg  position operator for the Hamiltonian $H_c$ : 
\begin {equation}
  x(t)=(\frac{p_xt}{m}+x_0) \label {CAC} 
\end {equation}
where $x_0=x(0)$ is Schrodinger position operator
If $\bar{x}_0=0$
the corresponding clock time operator, which will be extensively
used in relativistic theory can be decomposed as :
$$
    \hat{\tau}=t+\frac{p_x-\bar{p}_x+x_0m}{\bar{p}_x}
$$
The first term gives the time expectation value and the rest gives
the clocks dispersion $D(t)$. 
 To simplify our discussion  we'll consider 
 also the clocks model $C_0$ with
 the linear approximation of the position
operator  $x(t)=\omega t+x_0$ where  parameter $\omega=\frac{\bar{p}_x}{m}$
 which is the analog of Peres clocks for unbounded motion.
$C_0$ Hamiltonian $H_c^0=\omega p_x$ is unbounded 
from below for the continuous spectra, but for the interpretation
 of the relativistic clock effects it's unimportant.
 Any initial $C_0$ state ($\ref{CD}$) evolves as
 $\varphi^0_c (x-\omega t)$, so 
the initial form of wave function is conserved and only its centre
 of gravity moves.

 Now we'll consider the relativistic $C_x$ model in which
 RF $F^1$ and the particle $G^2$  system $S_2$ motion
is relativistic.
We'll suppose that ARF proper time $\tau_0$ is defined also by some
quantum clocks ,which dispersion is so small that can be neglected and
$\tau_0$ is the parameter. If 
$F^1$ internal interactions neglected 
 $F^1$ c.m. motion described by the massive boson wave packet evolution
 and $S_2$ Hamiltonian $H_T$
  in ARF is  the sum of two  
 KGR Hamiltonians for the positive energy states
 $\cite{Schw,Git2}$: 
\begin {equation}
%-i\frac{d \Psi_2}{d \tau_0}
H_T=(m_1^2+\vec{p}_{1}^{\,2})^{\frac{1}{2}}+(m_2^2+
\vec{p}_{2}^{\,2})^{\frac{1}{2}}=(s+\vec{p}_s^{\,2})^{\frac{1}{2}}
   \label {C0}
\end {equation}
, where $\vec{p}_s=\vec{p}_1+\vec{p}_2$ and $s$ is invariant mass square.
 $\sqrt{s}$ can be  regarded as
 the Hamiltonian of two objects $G^2, F^1$
 relative motion in their c.m.s. equal to system $S_2$ mass operator :
\begin {equation}
 m_t= \sqrt{s}=(m^2_1+\vec{q}^{\,2})^{\frac{1}{2}}
+(m^2_2+\vec{q}^{\,2})^{\frac{1}{2}} \label {C2XYY}
\end {equation}
 where $\vec{q}$ is $G^2$ relative invariant momentum \cite{Coe}. 
If $|\bar{q}|$ is small we can choose as $p_x$ - clock momentum $\vec{q}$ 
projection along any suitable direction for which $\bar{q}_x\ne 0$. 
 In this case $F^1 , G^2$ relative motion 
can be regarded as nonrelativistic and $F^1$ mass operator 
 approximated :
$$
 m_t\simeq m_s+\frac{q_x^2}{2\mu_{12}}+E_k(q_y,q_z)
$$
 ,where $\mu_{12}$ is $G^1,F^2$ reduced mass, $m_s=m_1+m_2$ is $S_2$ rest mass.
In this case $E_k$ is small  and can be omitted in the calculations. 
Like in  nonrelativistic case
 $F^1$ proper time in this $C_x$ relativistic model can be estimated
measuring in ARF the distance $x=x_2-x_1$ between $F^1$ and the particle
$G^2$ which operator is equal to :
$x=i\frac{\partial}{\partial q_x}$. For the obtained  $m_t$
 $S_2$  Hamiltonian $H_T$ can be  formally rewritten :
\begin {equation}
   {H}_T=[(m_s+H_c)^2+\vec{p}_s^{\,2}]^{\frac{1}{2}} \label {CA11}
\end {equation} 
 where $H_c=\frac{q_x^2}{2\mu_{12}}$. 
  Moreover it is reasonable to assume that this 
 square root  Hamitonian  can
describe the evolution of any clocks model with nonrelativistic interactions
 $H_c$ i.e. for $\alpha_I\ll 1$ $ \cite{Schw}$.
 Here and below the algebraic 
operations with the operators (if they don't result into singularities)
 means Tailor raw decomposition. If
$F^1, G^2$ relative motion is nonrelativistic we can assume for the beginning 
that $F^1$ and $S_2$ c.m.s. proper time practically coincide.
For the classical motion $F^1$ Lorentz factor in $S_2$ c.m.s.
$(1+\frac{\vec{q}^{\,2}}{m^2})^{\frac{1}{2}}$ and below we'll show that in quantum
case their difference is also negligible.
It's impossible to resolve in analytical form the Schrodinger
equation for $H_T$ of ($\ref {CA11}$) , only some approximated
solutions discussed below can be found.    
 $S_2$  observables evolution
  can be found solving Heisenberg
 equation for the Hamitonian $H_T$ of (\ref{CA11})
or for exact Hamitonian of (\ref{C0}) as will be done below   \cite {Hol}. 
 After the simple algebra one obtains  $x$ evolution
in  ARF proper time $\tau_0$ :
\begin {eqnarray}
  \dot{x}=-i[x,H_T]=
\frac{ -im_t}{(m_t^2+\vec{p}_s^2)^\frac{1}{2}}[x,H_c]=-iB_1[x,H_c]
  \label {C0H}
\end {eqnarray}
We'll call the operator $B_1(\vec{p}_s,m_t)$ the time boost operator,
 which interpretation will be discussed after some calculations.
 The clock observables  we obtain in  this clock models 
are the functions of canonical momentums only and due to it their
factor ordering is unimportant for our problem.
 After the commutators calculations we can approximate operator $m_t$ by
the parameter $m_t\simeq m_s+\frac{\bar{q}_x^2}{2m}$.
The operator $x$ easily restored from $\dot{x}$ :
$$
 x(\tau_0)=B_1(\vec{p}_s,m_t)\frac{q_x\tau_0}{\mu_{12}}+x_0 
$$
where  $x_0$ is Schroedinger position operator for $\tau_0=0$.
 If we take that $\bar{x}_0=0$ it results into $F^1$ proper  time operator :
\begin {equation}
\hat{\tau}_1= B_1(\vec{p}_s,m_t) \frac{q_x}{\bar{q}_x}\tau_0
+\frac{\mu_{12} x_0}{\bar{q}_x} \label {C2XX}
\end {equation}
Its meaning will be discussed after some calculations, but formally
it's $F^1$ moving clocks hand position  measured in ARF
at the moment $\tau_0$.
$\tau_1$ operator in $C_0$ model have the simpler form 
which prompts its interpretation  :
\begin {equation}
\hat{\tau}_1= B_1(\vec{p}_s,m_t) \tau_0
+\frac{ x'_0}{\omega} \label {CXT}
\end {equation}
If $\bar{x}'_0=0$ $C_0$ 
 $\hat{\tau}_1$ expectation value $\bar{\tau}_1=\bar{B}_1\tau_0$
coincides with the classical Lorentz time boost value.
 Its dispersion have the form : 
\begin {eqnarray}
D_{\tau}=D_L(\tau_0)+D_c=D_B\tau_0^2+\bar{D}_2\tau_0+D_0 \label {C2X}
\end {eqnarray}
where $D_B=\bar{B}^2_1-(\bar{B}_1)^2$
 and
 $D_0=\langle\frac{x_0^{'2}}{\omega^2}\rangle$ is the  clocks
mechanism dispersion, which for $C_0$ is time independent. 
 Operator $D_2$  is equal to :
\begin {equation}
D_2=\frac{B_1 x_0+x_0 B_1}{\omega}
  \label {C2ZZ}
\end {equation}
The numerical calculations show that for $C_0$ localized states
 $D_2$ expectation value is very small and  can be neglected.
If $D_0$ is  small $\tau_1$ fluctuations are defined mainly by  
$D_L(\tau_0)$ Lorentz boost  dispersion  stipulated by
 $\vec{p}_s$ fluctuations in $F^1$ wave packet.
It's independent of the clocks mechanism and demonstrates that
the proper time measurement have the principal quantum uncertainty
growing unrestrictedly proportional to $\tau_0^2$.
 
For $C_x$ model the factor $\frac{q_x}{\bar{q_x}}$ in ($\ref{C2XX}$)
produces additional 
$\hat{\tau}_1$ fluctuations.
Due to it Lorentz boost expectation value differs only for
 the small factor  of the order $\alpha_I$ :
$$
    \bar{\tau_1}=\tau_0\bar{B}_1[1+\frac{\bar{B}_1}{\sigma_x^2\mu_{12}m_s}
(1-\bar{B}_1^2)]
$$
It results from  $m_t$  dependence on $p_x$ and reflects influence
 of clocks energy on total mass. We'll neglect this effect in $C_x$ dispersion
 also described by ansatz (\ref {C2X}),
 but with different  parameters :
\begin {eqnarray}
D_2=\frac{\mu_{12}}{\bar{q}_x^2}(q_xB_1x_0+x_0q_xB_1) ; \\
D_B=\frac{\bar{q}_x^2}{(\bar{q}_x)^2}\bar{B}_1^2-(\bar{B}_1)^2 
\label {C2Y} ; \quad
D_0=\frac{ \mu_{12}^2 \sigma_x^2}{\bar{q}_x^2} \label {C2WW}
\end {eqnarray}
Here $\bar{D}_2=0$ for the gaussian wave packets (\ref{CD}) 
and any other localizable states. 
Due to  $q_x$ fluctuations absent in $C_0$ model  
the part of $D(\tau_1)$ :
$$
D_x= D_0
 +\frac{\bar{q}_x^2-(\bar{q}_x)^2}{(\bar{q}_x)^2}(\bar{B}_1)^2\tau_0^2
$$
can be related to the packet smearing along $x$ coordinate, 
regarded as the clocks mechanism uncertainty. 

To illustrate the physical meaning of this time operator
 let's consider the corresponding approximate solutions of $F^1$ state
  evolution equation for Hamiltonian  (\ref{CA11}).
For $\alpha_I\rightarrow 0$ we can decompose $H_T$ of (\ref{CA11}) in the
first  $\alpha_I$ order :
\begin {equation}
-i\frac{d\Psi_s}{d\tau_0}
%      =(m_s^2+\vec{p}_s^2)^{\frac{1}{2}}\Psi_2
 \simeq
[(m_s^2+\vec{p}^2_s)^{\frac{1}{2}}+
\frac{m_1\hat{H}_c}{(m^{2}_s+\vec{p}_s^2)^{\frac{1}{2}}}]\Psi_s
                \label {C01}
\end {equation}
Here the first term is independent of $H_c$
which permit to represent $\Psi_s$ as the sum of factorized states.
The second term is in fact the product of clock Hamitonian and
Lorentz boost $B_1$.
 Let's choose the initial $F^1$ state
 $\Psi_s(0)=\Phi_s(\vec{p}_s)\varphi_c(x,0)$ and
% $\varphi_c=|v_0\rangle$
 $\Phi_s=\sum c_l|\vec{p}_{sl}\rangle$, where the sum 
denotes the integral over $\vec{p}_s$. From our definition of
quantum clocks synchronization it follows that
 $\Psi_s(0)$ describes $F^1$ clocks
synchronized with ARF clocks at $\tau_0=0$. Solving equation ($\ref{C01}$)
  one finds  : 
\begin {equation}
\Psi_s(\tau_0)=\sum c_l\varphi_c(x ,B_l\tau_0)
|\vec{p}_{sl}\rangle e^{-iE(\vec{p}_{sl})\tau_0} \label {C0A}
\end {equation}
%where $\varphi_{2l}(u_i,0)=\varphi_2(u_i,0)$,
where $E(\vec{p})=(m_s^{2}+\vec{p}^2)^\frac{1}{2}$ ,
$ B_l=B_1(\vec{p}_{sl},m_s)$.
For linear clock $C_0$ Hamiltonian $H_c=H_c^0$ 
  for small $\alpha_I$  this state can be rewritten :
\begin {equation}
\Psi_s(\tau_0)=\sum_l c_l \varphi^0_c(x-\omega B_l\tau_0)
~|\vec{p}_{sl}\rangle e^{-iE(\vec{p}_{sl})\tau_0} \label {C0B}
\end {equation}
To make the situation more clear  suppose that
 $\varphi^0_c(0)=\delta(x)$,
which evolves at rest into $\delta(x-\omega \tau_0)$  .
 Then $x$ measurement defines the time $\tau$ of quantum clocks at rest
 unambiguously and with zero dispersion,
 but  $\Psi_s$ of ($\ref {C0B}$) in general isn't
$x$ eigenstate.   It means that at any $\tau_0>0$  
$\Psi_s$ is the entangled superposition of the states $\varphi_c^0$
 which $F^1$ clocks acquires at the consequent $\tau_1$ moments.
As was shown there is one-to one correspondence between clock state 
$\varphi_c(x,t)$ and the time moment $t$ and in some sense
it can be regarded as the 'superposition' of $F^1$ proper time moments, or more
precisely $F^1$ states existed at this moments.
 For example   $F^1$ clocks hand
 can show 3,4 and 5 o'clocks simultaneously 
which can be tested by $x$ measurement at some $\tau_0$ in ARF.
This spread corresponds to $D_B$ dispersion term resulting from the
$F^1$ momentum $\vec{p}_s$ uncertainty. For the realistic clocks their
$x$ dispersion  given by $D_0$ isn't zero even at rest and this
two terms added as statistically independent effects.
$\Psi_s$ for $C_x$ Hamitonian is given by ($\ref{C0A}$) and
admits the same interpretation. It corresponds to the more  
complicated form of time dependent dispersion ($\ref{C2WW}$) which
can be eventually factorized into the same two parts -
relativistic and  clock mechanism. 
So we conclude that the interpretation which follows from the
approximate Schrodinger equation agrees well with Heisenberg
operator calculus. In fact operator $\tau_1$ describes $F^1$
proper time in the limit when this clock dispersion is very
small and the clock energy is much less then $F^1$ total mass
 energy i.e. $\alpha_I\rightarrow 0$.

Obtained results suppose that the proper time of any quantum RF
 being the parameter in it simultaneously
 will be the operator from the 'point of view' of  other RF.
Qualitatively the appearance of RF proper time fluctuations
can be understood considering the superposition of momentum
eigenstates $|\vec{p}_{si}\rangle$ in $S_2$ wave packet 
as the superposition of $S_2$ velocities $\vec{\beta}_i$ and  
corresponding Lorentz factors $\gamma_1(\vec{\beta}_i)$. 
In Special Relativity 
$F^1$ proper time $\tau_1$ measured at the same $\tau_0$ in ARF
depends on $\gamma_1$.  If we formally extends this dependence
  on $F^1$ wave packet motion we get that the  proper time
  will fluctuate proportionally to  $ \gamma_1$ spread.
So $F^1$ clocks measurement in ARF shows how much time passed in $F^1$ 
in this particular event and can give the different value
for another event of the same ensemble.    
It means that the time moments in different RFs corresponds only
statistically with the dispersion $D_{\tau}$ in ARF 
given by (\ref {C2X}). It differs from Special Relativity
where one to one correspondence between $\tau_1, \tau_0$ 
time moments always exists , but can be incorporated into
relativistic QEP if we find the analogous time relations
between two quantum RFs of finite mass.

In fact $\tau_1$ is more correct to relate to $S_2$ c.m.s. rest frame, but
regarding the difference between $F^1$ and $S_2$ c.m.s. proper time
operators $\tau'_1,\tau_1$ it's easy to show that they coincide
if $\bar{q}_x\rightarrow 0$. From it 
we  conclude that the principal part of the relativistic time operator,
 independent of any particular clocks mechanism features have the form in
the limit $\alpha_I\rightarrow 0$ :
\begin {equation}
\hat{\tau}'_1=B_1(\vec{p}_1,m_1)\tau_0   \label {CBBB}
\end {equation}
 Moreover this formulae permits
to define formally the time operator for any object including
the single massive particle.
 This operator form of $\tau'_1$ is closely
connected with Fock-Shwinger proper time $\tau_F$ formalism
 interpretation and will be discussed in detail in the forcoming paper
 $\cite{Fock,Schw}$. Note only
 that $\hat{\tau}'_1(\tau_0)$ measurement gives $F^1$ proper time
 $\tau_F$ estimate
 at $\tau_0$ moment of ARF time. On the opposite in  Fock-Shwinger formalism 
$\tau_F$ is the parameter  time to which particular values operators
 $\hat{\tau}_0(\tau_F), \vec{r}_1(\tau_F)$ related. In distinction with our
formalism it makes $\tau_F$ interpretation confusing, because
$\vec{r}_1$ and other $ F^1$ operators are measured in ARF, hence
the time of measurement defined in $F^1$ to which as we have shown
 in quantum case they related only statistically.

 The practical realization of  $x$
 measurement in ARF can be the intricated procedure, which
scheme we don't intend to discuss here.
Note only that to perform it one should measure simultaneously
the distance between $F^1$ and $G^2$ and their total momentum
 giving total velocity and this two operators commute.
 Some examples of the analogous
nonlocal observables measurements are described in \cite{Aha2}.
The most disputable question here is the relativistic particle
coordinate measurements. Yet in the considered case, when the
relative $F^1,G^2$ average velocity  is small
 then $x$ is  the nonrelativistic coordinate operator. Yet to prove
 the quantum equivalence principle it's necessary to perform
the full relativistic calculations. We'll present such completely
relativistic results for $C_x$ model using
 Newton-Wigner Hermitian operator of the space coordinate 
  $\cite{Wig}$ which is the direct analog of nonrelativistic operator $x_1$ :
\begin {equation}
\hat{x}^1_{NW}=i\frac{d}{dp_{x1}}
-i\frac{p_{x1}}{2(m_1^2+\vec{p}^2_1)} \label {C5}
\end {equation}
 The operator of two objects relative coordinates
 conjugated to c.m. momentum $q_x$
 can be derived from this objects
c.m. Hamiltonian (\ref{C2XYY}) :
\begin {equation}
\hat{x}_{NW}=x+F(\vec{q})=i\frac{d}{dq_{x}}
-i\frac{q_{x}}{\sqrt{s}}(\frac{1}{w_1}+\frac{1}{w_2})
%-i\frac{q_{x}(w_1^3+w_2^3)}{2w_1^2w_2^2(w_1+w_2)} \label {C5A}
\end {equation}
where  $w_i=(m_i^2+\vec{q}^2)^{\frac{1}{2}}$.
% $w^{-1}=(m_1^2+\vec{q}^2)^{-\frac{1}{2}}+(m_2^2+\vec{q}^2)^{-\frac{1}{2}}$.
The clocks time observable in $F^1$ rest frame is  proportional to $x_{NW}$ :
$$ 
   \tau=\frac{x_{NW} - \bar{x}_{NW}(0)}{\bar{\beta}_x}
$$   
where $\beta_x=q_x(w_1^{-1}+w_2^{-1})$ is  $F^2,G^1$
relative velocity, 

If we choose $\bar{x}_{NW}(0)=0$ , then solving
 Heisenberg equation in ARF for the Hamiltonian of
 (\ref{C0})  we find  the resulting $F^1$ time operator  :
\begin {equation}
   \hat{\tau}_1=  \frac{B_1(\vec{p}_s,m_t){\tau_0}\beta_x+x_{NW}(0)}
{\bar{\beta}_x}
\label {C5AA}
\end {equation}
where in $B_1$   $m_t=\sqrt{s}$.
This is the exact relativistic expression for $\tau_1$ 
without assumption of $q_x$ smallness. 
$\bar{\tau}_1$ corresponds to Lorentz boost value $\bar{B}_1$ which depends 
both on 
$\langle \vec{p}_s \rangle$ and $\langle \vec{q} \rangle$.
It's easy to note that the momentum dependent
part of $x_{NW}$ is constant in time and consequently can only
enlarge the clocks mechanism dispersion $D_0$.
Due to it the dispersion structure is the same as for nonrelativistic
relative motion of (\ref{C2WW}) but its members are
 described by the more complicated formulaes omitted here.
 In fact this calculations evidence 
that $x_{NW}$ meausurements introduces only additional
time-independent clock dispersion of the
order of $G^2$ Compton wavelength
without changing our previous conclusions about time operator properties. 

In fact $F^1$ proper time  measurement in ARF can be performed by
two different methods which equivalence must be proved.
In the first method described above the  detector $D_0$ installed in ARF
measures $\tau_1$ and induces $C_x$ state collapse. 
In the second one the detector $D_1$ installed in
 $F^1$  measures the clock state and after it $D_1$ signal
transfered to ARF. In this case we should consider the collapse
 in the moving frame , which is difficult to describe.  But we must
note that independently of its mechanism  such interaction happens 
after this clocks evolves to this state and so can't influence
directly on their evolution, so it seems correct to neglect it
at this stage.
Obtained time-fluctuation effect reminds the well-known life-time dilatation
 for the relativistic 
unstable particles $\cite{Byc}$. In this framework such particle
 can be regarded as the elementary  binary clock having only two states.

 Obtained results evidence that the proper time in Quantum RF depend
on the  RF quantum state, but doesn't prove QEP directly. 
To do it we must consider two  finite mass RFs on equal ground 
and to find the time transformation between them.

\section {Relativistic Quantum Frames}

To calculate the time operator between two RFs of finite mass
 it's necessary first to find the particle  evolution equation in
quantum RF rest frame. In general the system Poincare group
 irreducible representations contain the information which permit
 to describe its evolution completely, but 
due to appearance of time operators to find this representations
for quantum RF is quite a problem. Therefore we choose another route;
 first we'll find the free particle evolution equation and corresponding
proper time operator from
Dirack constraints quantization. After it we'll investigate
 Poincare  transformations  for quantum RFs with the clues prompted by 
 this Hamiltonian ansatz.

Dirack constraint formalism which permit to define 
free particle/antiparticle positive Hamiltonian was developed by Gitman
and Tiutin $\cite {Git2}$. They've shown that starting from
free scalar particle action $S=-\int mds$ Dirack constraint quantization
of $p_{\mu}^2-m^2$ initial superhamiltonian
results  into positive square root Hamiltonian $H_p$ as
function of 3-momentum $\vec{p}$ plus additional
charge $\xi=\pm 1$ discriminating antiparticles. In quantum case
it was shown to result in Klein-Gordon square root (KGR) equation for
both $\xi$
$$ 
  -i\frac{d\psi}{d\tau_0}=\sqrt{\vec{p}^2+m^2}\psi
$$

 Following this approach
we start from classical two particles action defined in ARF $F^0$ with
time parameter $\tau_0$ :
$$
   S(\tau_0)=\int L d\tau_0=   \\
 \int m_1[(\dot{x}^2_{10}-\dot{x}^2_{1i})^{\frac{1}{2}}
+m_2(\dot{x}^2_{20}-\dot{x}^2_{2i})^{\frac{1}{2}}] d\tau_0
$$
From it one finds 4-momentums $\pi_{1\mu}, \pi_{2\mu}$
which satisfy to superhamiltonian constraints $\pi_{j\mu}^2=m^2_j$.
%Gitman and Tuitin resolved this constraint for single
%free particle reducing its degrees of freedom to 3-momentum and charge $\xi$.
Due to $m_1, m_2$ dynamics independence $S_2$ system Dirack quantization
results in double number of 3- momentums $\vec{p}_{j}$ and charges
$\xi_{j}$. After simple calculations repeating
Gitman-Tiutin anzats one obtains system Hamiltonian :
$$
    H_T=\sqrt{\vec{p}_1^2+m_1^2}+\sqrt{\vec{p}^2_2+m_2^2}
$$         
For two particles system quantization we can use this Hamiltonian
in Schrodinger equation. But if we consider $m_1$ as quantum RF
then like in nonrelativistic case, before  quantize it we must 
put additional constraints corresponding to
  $F^1$ choice as rest frame
and defining $m_2$ operators transformations into it.
 Analogously to nonrelativistic case we put
constraint on $F^1$ momentum in its rest frame $\vec{p}_{11}^2\approx 0$,
meaning that RF don't move relative to itself. 
For charges we put  constraint $\xi'_1=\xi^2_1$ - .i.e. RF can't
 be antiparticle for itself, and correspondingly $\xi'_2=\xi_1 \xi_2$. 
Below for simplicity  we'll describe in detail only restricted  Hilbert space 
 sector without antiparticles  $\cite {Git2}$.
Corresponding to $H_T$ two particles state vector
isn't  interpereted by us as their wave function in $F^0$. For two particles
correlations it can result into contradictions connected with
nonlocalities. But we don't need  it in such role and
will study only reduced state vector in quantum RF $F^1$.

 To find the transformations from ARF to $F^1$
 we consider the system $S_2$ of RF $F^1$   and   particle $G^2$
 which momentums $\vec{p}_i$, energies $E_i$ are defined in  ARF.
% $S_2$   Lorentz generators  are $P_{s\mu} , \vec{N}_s,\vec{J}_s$.
  For the constraint described above we choose
 ARF FC $\vec{p}^{\,2}_0\approx 0$ and $S_2$ Hamitonian :
\begin {eqnarray}
  H^0_A=(m_0^2+\vec{p}_0^2)^\frac{1}{2}+
 (m_1^2+\vec{p}^2_{1})^\frac{1}{2}+
 (m_2^2+\vec{p}^2_{2})^\frac{1}{2}  \label {C2}
\end {eqnarray} 
if $G_2$ is boson. Like in nonrelativistic case 
all $\vec{p}_i$ are the operators and
  state vector  ascribed also to ARF $|\vec{p}_0=0\rangle$. This
ARF constraint formalism 
reproduces all relativistic QM results for $m_0\rightarrow\infty$.
Performing transformations  to $F^1$ rest frame we assume
 that the proper time parameter 
$\tau_1$ can be defined in it from $F^1$ clocks measurements extrapolation
as was described in previous chapter.
% Choosing $F^1$ FC $\vec{p}^{\,2}_{11}\approx 0$  from the correspondence
% principle we'll suppose  that the  momentums expectation values
%  in $F^1$ rest frame are given by Lorentz transformations with velocity 
Then from $\vec{p}_1$ constraint and correspondence with
 Lorentz  momentum transformations 
we phenomenologically find  $m_i$ momentums:
\begin {eqnarray}
  \vec{p}_{12}=\vec{p}_i+
\frac{(\vec{n}_1\vec{p}_i)(E_1-m_1)\vec{n}_1-E_i\vec{p}_1}{m_1} 
  +\vec{F}_i(\vec{p}_0)  \label {C20}
\end {eqnarray}
where $\vec{n}_1=\vec{p}_1 |\vec{p}_1|^{-1}$.
$\vec{F}_i$  are undefined at this stage operators for which 
$\langle \vec{F}_{i}\rangle=0$ and  can be neglected
 in the following calculations.
This transformation 
results in $\vec{p}_{11}=0$ if $\vec{p}_0=0$ and will  advocate its form
below where we'll discuss Poincare group for quantum RFs.
Until  then this is phenomenological transformation which for definite
RFs momentum and velocities reproduces Lorentz transformations.
 If $G^2$ have spin zero then the Hamiltonian
$H$ transformed from ARF to $F^1$ is equal :
\begin {eqnarray}
  H^1_{Tot}=H^1_0+H^1_1+H^1_2=
\sum_{i=0}^2 (m^2_i+\vec{p}^{\,2}_{1i})^\frac{1}{2}  \label {CC2}
\end {eqnarray}
For classical Special Relativity where normally RF 
supposed to  have infinite mass  $\vec{p}_{1i}, H^1_R$ corresponds
to the  canonical momentums
for finite mass RFs $\cite {Lan}$. We see that
$F^0$ motion Hamiltonian is factorized and so we can drop it
 and  regarding $S_2$ motion can use  $S_2$ Hamiltonian
 $H^1=H^1_1+H^1_2$. In quantum case in $H^1$ we can't simply omit 
 $\vec{p}_{11}$ because now it's operator.
So in $F^1$ proper time $\tau_1$ $S_2$ evolution equation is :
\begin {equation} 
 -i\frac{d\psi^1}{d\tau_1}=
[(m_1^2+\vec{p}_{11}^{\,2})^\frac{1}{2}+
(m_2^2+\vec{p}^{\,2}_{12})^\frac{1}{2}]\psi^1
=(s(\vec{q})+\vec{p}_s^{\,2})^\frac{1}{2}\psi^1
 \label {C4}
\end {equation}
where  $S_2$ c.m. observables $\vec{q},\vec{p}_s$ defined in chap.3.
Solutions of this equation describe $G^2$ normalized
free wave packet localizable relative to $F^1$ rest frame : 
\begin {equation}
  \Psi^1(\tau_1)=\varphi_2(\vec{p}_{12}) e^{-iE^1\tau_1}
|m_2 ,\vec{p}_{12}\rangle
|m_1,\vec{p}_{11}=0\rangle=\\
\varphi'_2(\vec{q}) e^{-iE^1\tau_1}
|\sqrt{s} ,\vec{p}_s=\vec{p}_{12}\rangle
|m_1,-\vec{q}\rangle|m_2,\vec{q}\rangle
   \label {CD1}
\end {equation}
expressed  also via $S_2$ c.m. observables.
Here $E^1=E^1_1+E^1_2$ are $H^1$ eigenvalues. They differ from the standard 
KGR energy only on $m_1$ and so
  we can use in $F^1$ rest frame the standard KGR  momentum  spectral 
decomposition and the states scalar product $\cite{Schw}$. 

In $F^1$ rest frame together with its proper time
$\tau_1$  the space coordinate can be defined. We choose
arbitrarily as $G^2$  coordinate  (nonhermitian) operator in $F^1$  :
$\hat{x}_{12}=i\frac{\partial}{\partial q_{x}}$ and
corresponding Hermitian Newton-Wigner operator can be easily derived.
Note that $x_q$ defined in $F^1$ differs from the same operator 
defined in c.m.s.,
yet our following results doesn't depend on the particular form
of this operator.  $x_{12}$ also differs from the operator
$x_{p}=i\frac{\partial}{\partial p_{12x}}$ which corresponds to the
classical distance between $F^1$ and $G^2$. They coincide only
in the limit $m_1\rightarrow\infty$ or in nonrelativistic case.

Now we can calculate  
 $F^2$ proper time operator  as  function of the proper time in $F^1$.
To perform it we assume again that $F^2$ c.m. motion is equivalent to 
the spinless particle $G^2$ motion.
 In the described framework the Hamiltonian of $F^2$ with $C_0$ or $C_x$
 clocks in $F^1$ rest frame can be obtained substituting
in $\hat{H}^1$ of (\ref{C4}) $m_2=m'_{2}+\hat{H}_c$.
  $\hat{\tau}_2$ can be found
solving Heisenberg equation for $F^2$ clocks coordinate 
$\dot{x}=-i[x,H^1]$ analogously to  ($\ref{C0H}$).
  If we omit 
analogously to (\ref{CBBB}) the members describing the clocks mechanism
 fluctuations
 the $F^2$ proper time operator
 $\hat{\tau}_2$  is equal  :  
\begin {equation} 
    \hat{\tau}_2=\frac{m_2 \tau_1}{(m^2_2+\vec{p}_{12}^2)^\frac{1}{2}}
\approx \frac{m'_2 \tau_1}{(m'^2_2+\vec{p}_{12}^2)^\frac{1}{2}}
=\hat{B}_1 (\vec{p}_{12},m'_2)\tau_1  \label {C6}
\end {equation}
 This formalism is completely symmetrical and the
 operator obtained from (\ref{C6}) exchanging indexes 1 and 2
relates the
 time $\hat{\tau_1}$ in $F^1$ and $F^2$ proper time
- parameter $\tau_2$.
The Special Relativity limit when $\tau_2$ becomes
 the parameter is obvious 
and analoguosly to it the average time boost depends on whether
 $F^1$ measures $F^2$ clocks observables, as we consider or vice versa,
 and  this measurement  makes  $F^1$ and $F^2$ 
nonequivalent $\cite{Lan}$.  The  new effect will be found only
when $F^1$ and $F^2$ will compare their initially synchronized clocks.
In QM formalism this synchronization means that
 $F^2$ state prepared at the moment $\tau_0$ can be factorized as
 $\Phi_2(\vec{p}_{12}) \varphi_c(x,0)$ analogous to ($\ref{C0A}$).
If this $F^2$ time measurements repeated several times
(to perform quantum ensemble) it'll reveal not only 
classical Lorentz  time boost ,
 but also the statistical spread having quantum origin with the
dispersion given in (\ref{C2X}). Obtained relation between  two 
finite mass RFs proper times evidence that Quantum Equivalence principle
can be correct also in relativistic case.

If the number of particles $N_g>1$ then for the system state description   
the clasterization formalism can be used 
    $\cite{Coe}$. According to it for $N=3$
 Hamiltonian in $F^1$ of two free particles $G^2,G^3$ rewritten through
the  system canonical observables acquires the form   :
\begin {equation}
   \hat{H^1}=
   (m_1^2+\vec{p}_{11}^2)^\frac{1}{2}+(s_{23}+\vec{p}^2_{1,23})^\frac{1}{2}
=(s+\vec{p}_s^2)^\frac{1}{2} \label {C7}
\end {equation}
,where $\sqrt{s}_{23}$ is $G^2, G^3$ 
invariant mass, $\sqrt{s},\vec{p}_s$ are the system total invariant mass and
 momentum. In clasterization
formalism at the first level the relative motion of $G^2, G^3$ defined by
$\vec{q}_{23}$ their relative momentum is considered.
 At the second level we regard them as the single quasiparticle
 - cluster $C_{23}$ with mass $\sqrt{s}_{23}$ and momentum $\vec{q}$
in the system c.m.s. It transformed to
  $\vec{p}_{1,23}$   momentum in $F^1$
and  so  at any level we can regard 
 the relative motion of two objects only. This
 procedure can be extended in the obvious inductive way to  
  arbitrary $N$. If we have two reference frames $F^1,F^2$ and $N_g\ne0$
 then their relative momentums can be also described by the cluster formalism.

Due to appearance of the time operator between two RFs 
 to find Poincare group  transformations for quantum RFs
$\hat{U}^s_{2,1}(\tau_2,\tau_1)$ is quite a problem and
here we can present it only phenomenologically  for some simple examples.
We don't include rotations into consideration, so this results are 
suitable completely only for 1-dimensional case and for more
dimensions they give only partial description of the Lorentz transformations.
 Consider first the case $N=2$ when $S_2$ include $F^1,F^2$ only and 
 its  state in $F^1$ rest frame 
$\Psi^1(\tau_1)$  is the solution (\ref{CD1}) of eq. (\ref{C4}).
We'll take that it transformed by $U^F_{2,1}$ into  state $\Psi^2(\tau_2)$
 in $F^2$ rest frame. 
If $F^1,F^2$ clocks are synchronized at $\tau_1=\tau_2=0$ then for
this time  moment
  $\Psi^2(0)=\hat{U}^F_{2,1}(0,0)\Psi^1(0)$ and from $F^1, F^2$ symmetry 
it follows :
 $|\Psi^2(0)\rangle=\varphi'_1(\vec{p}_{21})
|m_2,\vec{p}_{22}=0\rangle|m_1,\vec{p}_{21}\rangle$.
$F^{1,2}$ internal wave functions $\varphi^{1,2}_c(x,0)$ at $\tau_1=0$
 are obviously invariant and so omitted here.
 Like in nonrelativistic case we introduce
 $\vec{p}_f=\vec{p}_{11}+\vec{p}_{12}$, 
$\vec{p}'_f=\vec{p}_{21}+\vec{p}_{22}$ and conjugated $\vec{r}_f, \vec{r}'_f$.
From the correspondence with Lorentz transformations it should give
$\langle\vec{p}_{12}\rangle=-\frac{m_2}{m_1}\langle\vec{p}_{21}\rangle$
and if to demand  that fro relative momentum in $F^2$ 
$\vec{q}_2=-\vec{q}$ must be fulfilled
then  the simplest transformation is  :
\begin {equation}
     \hat{ U}^F_{2,1}(0,0)=P_f e^{i a_f\vec{p}_f\vec{r}_f}
   \label{CBBW}
\end {equation}
where $a_f=\ln\frac{{m}_1}{m_2}$,
% \vec{b}_s=\frac{M}{m_0}\rangle\vec{R}_c\langle$.
  $P_f$ is $\vec{r}_f$ reflection (parity) operator. 
We see $\hat{U}^F_{21}(0,0)$ ansatz practically coincides with nonrelativistic
transform of (\ref{BBB}) for $N=1$.
The passive $S_N$  transformation for spinless $G^i$ also found
from the correspondance principle as the minimal extension of
standard Poincare transformations : 
\begin {equation}
\hat{U}^s_{2,1}(0,0)=U^F_{21}(0,0)\prod_{j=3}^N
 e^{-i\vec{\beta}_{f}\vec{N}'_j} \label {C08}
\end {equation}
 where  velocity operator $\vec{\beta}_{f}=\vec{p}_{f}(H^{1}_2)^{-1}$,
 $\vec{N}'_i=H^1_{i}\frac{\partial}{\partial\vec{p}_{1i}}
+\frac{\partial}{\partial\vec{p}_{1i}}H^1_i$
 are $G^i$ Poincare generators in $F^1$ which coincide with standard ansatz.
 Then the transformation operator for arbitrary $\tau_1,\tau_2$ is :
 \begin {equation}
\hat{U}^s_{21}(\tau_1,\tau_2)=
\hat{W}_2(\tau_2)\hat{U}^s_{21}(0,0)\hat{W}_1^{-1}(\tau_1) \label{C8}
\end {equation}
, where  $\hat{W}_{1,2}(\tau_{1,2})=exp(-i\tau_{1,2}\hat{H}^{1,2})$
are $S_N$ evolution operators and $H^{1,2}$
- $S_N$ Hamiltonians in $F^1,F^2$ rest frames. 

 It means that despite $\tau_2$
and $\tau_1$ are correlated only statistically through $\hat{\tau}_2$
 nevertheless  $S_N$ state vectors for free motion in 
$F^2, F^1$  at this moments are related unambiguously.
Transformed $S_N$ momentums are :
\begin {eqnarray}
\vec{p}_{21}=-\frac{m_2}{m_1}\vec{p}_{12}+d_1\vec{p}_{11}  \quad ,
\vec{p}_{22}=-\vec{p}_{11}+d_2\vec{p}_{11}\\
  \vec{p}_{2i}=\vec{p}_{1i}+
\frac{(\vec{n}_{12}\vec{p}_{1i})(E^1_2-m_2)\vec{n}_{12}-E^1_i\vec{p}_{12}}
{m_2}   +d_i\vec{p}_{11}  \label {C99}
\end {eqnarray}
, where $\vec{n}_{12}=\frac{\vec{p}_{12}}{|p_{12}|}$, $E^1_i$
are $G^i$ energies in $F^1$.
 If to demand that all relative momentums $\vec{q}_{ij}$ conserved
(or reflected), then
  $d_i$  can be calculated, but due to their unimportance
 we omit it here. It's easy to see that $\vec{p}_{1i}$ of (\ref {C20})
for ARF to $F^1$ transform follows from $U^s_{2,1}$ after the simple
 substitutions, and so the semiqualitative Hamiltonian derivation
 of (\ref{C20}) was consistent. We see that the passive spinless $G^3$
  transformation differs from the standard one only by the change of
 velocity  parameter to the operator
$\vec{\beta}$ which commutes with  $G^3$ Hamiltonian.
  
It was argued that RF quantum properties can become important in Quantum
Gravity , where in principle one should quantize the field, matter
 and RF simultaneously
$\cite{Rov,Unr,Bro}$. In principle our approach permits to calculate 
the time operator $\hat{\tau}_1$ for RF $F^1$  moving in the external 
gravitational field $g_{\mu\nu}(x)$. We assume that ARF is located in
the region where this field is weak and so we can take $\tau_0=x_0$ -
world time. Analogously to (\ref{CA11}) $F^1$ clocks Hamiltonian in
ARF (for $g_{oa}=0$ gauge)  :
\begin {equation}
\hat{H}_T=[g_{00}(m_1+H_c)^2+g_{00}g_{ab}p^a_1p^b_1]
^\frac{1}{2}   \label{CG}
\end {equation}  
,where $a,b=1,3$  $\cite{Lan}$.
Now $H_T$ depends on $x_{\mu}$ and due to it
solving Heisenberg equation (\ref{C0H}) for the
 clocks hand coordinate $x_c$  
one obtains the differential relation  for $\tau_1$ :
\begin {equation}
d\hat{\tau}_1=\frac{\sqrt{g_{00}}(m_1+H_c)d\tau_0}
{[(m_1+H_c)^2+g_{ab}p^a_1p^b_1]
^\frac{1}{2}}=\sqrt{g_{00}}B_g(x,\vec{p}_1)d\tau_0           \label {CGG}
\end{equation}
In this case $\hat{\tau}_1$ becomes the integral operator , where
integral is taken over $\tau_0$ interval. If $g_{\mu\nu}$ is
 the classical metrics
then this relation contains no new physics , except the additional 
gravitational 'red shift' time boost proportional to $\sqrt{g_{00}}$
 $ \cite{Lan}$.
But in Quantum Gravity $g_{\mu\nu}(x)$ becomes the operator and its
fluctuations can induce the additional quantum fluctuations of the measured 
$F^1$ clocks time. Despite that this fluctuation calculations are
quite complicated we can expect from the general Quantum Statistics
rules $\cite{Hol}$ that they can be factorized from the
 considered Lorentz boost
fluctuations induced by the $F^1$ momentum fluctuations :
$$
   D_T=D_G(\tau_0)+D_L(\tau_0)+D_o(\tau_0)
$$
From this rules we can expect also that for $F^1$ motion
in the homogeneous gravitation field
$D_G$ will grows proportionally to $\tau_0$ analogous to QED fluctuations 
(Brownian motion effects). Note that this fluctuations
must be independent of RF mass.
  
This approach can give some new insight into the famous time problem of 
Quantum Gravity $\cite{Rov,Bro}$ which we discuss here briefly.
In this aspect the situation in Classical and Quantum Gravity
seems to differ principally.
 Strictly speaking if the metrics becomes the operator it stops to be
space-time metrics which unambiguously defines the space-time geometry.
Due to it the observer can correctly use only the operational
definition of physical space-time by means of clocks and other measurements.
In gravity this operational time can originate from
some evolving observable of gravitation field or
   to be the operator describing the time measurement
 for some free  matter object carrying some nongravitational
 'foreign' clocks.
The idea that space-time events can be described by their relation
with some distributed  system or media was extensively explored for long time
$\cite {Dew}$  .
The most close to our purposes is the incoherent dust system, each
piece of it carrying clocks. Gravity ADM quantization for
 such system with selfgravitation account permits
to extract positive Schrodinger hamiltonian as was shown by Brown and 
Kuchar $\cite{Bro}$. Hence the dust pieces motion  in their model was described
 only semiclassically. Introduction of  'dust space' $\vec{z}$ permit to 
 quantize the gravitation field. Yet the 
 free quantum motion of dust pieces transforms
$\vec{z}$  into the operator on the initial space-time
 manifold $x_{\mu}$ which makes this quantization procedure contradictory.

We describe here briefly
 the  model of dust RFs quantum motion where in the first approximatio
 its selfgravitation neglected. 
Let's consider first classical RF $F^1$ free falling in 
 external gravitational field. 
In $F^1$ comoving 'Gaussian' frame where frame conditions
imposed before the field variation we have $g'_{00}=1, g'_{0a}=0$.
In this RF  for the classical field gravity constraints
fulfilled $H_a(x)=0, H_0(x)=0$ which permit to calculate 
$g'_{ab},p'_{ab}$  evolution for  $F^1$ clock time solving corresponding
 Hamilton equations for $H_0$ $\cite {Bro}$. In quantum case
 this vacuum field constraints
results in Wheeler - deWitt equation $\hat{H}_0\Psi=0$ from which
Schrodinger Hamiltonian can't be derived easily. Now let's account
 RF quantum motion
 and  suppose that this constraints holds true  also in quantum $F^1$ 
 comoving frame. $F^1$ proper time for the external
observer  is given by the operator analogous to ($\ref{CGG}$),
but in comoving frame $\tau_1$ is just the parameter.
 In this case we can calculate
field observables evolution in $F^1$ clocks time 
 from  Heisenberg equations for $H_0$ vacuum constraint  :
$$
  \dot{g}'_{ab}(x)=-i[g'_{ab}(x),H_0(x)]
$$
where the commutator in general is nonzero. Note that
this  equation is obviously local, so to calculate $g'_{ab}(x,\tau_1)$
we must define $g'_{ab},p'_{ab}$ only on a small spacelike  surface region
around $x$ at a preceding moment $\tau_1-d\tau_1$. Space coordinates
$x_a$ supposedly can be defined at least in the close
vicinity of $F^1$ analoguosly to the definition given above for
the flat space-time.  
Obviously this approach have many associated problems some of which are the
 construction of multifingered time  for quantum RF dust
and  the field theoretical behavior of such commutators,
 despite it seems to deserve additional study. 

 For the conclusion we can claim that the extrapolation of QM laws on free
macroscopic objects regarded as RFs prompt to change
 the common approach to the 
space-time  which was taken copiously from  Classical 
Physics. In this paper the relativistic covariant theory
of quantum RFs constructed and at least in flat space-time
it agrees with the principle of equivalence for quantum RFs.
The quantum RF  momentum uncertainty results 
in the quantum statistical fluctuations of Lorentz boost which relates
the proper times in two RFs.
So in this model each observer has its proper time - parameter and
euclidian coordinate space which can't be   
related unambiguously with the another observers space-time
 and in this sense is local. 

\end{sloppypar}
{\footnotesize


\begin{thebibliography}{99}

\bibitem{Aha} Y.Aharonov, T.Kaufherr Phys. Rev. D30 ,368 (1984)
 
\bibitem{Dop} S.Doplicher et. al. Comm. Math. Phys. 172,187 (1995) 

\bibitem{Schiff} L.I.Schiff, 'Quantum Mechanics' (New-York, Macgraw-Hill,1955) 

\bibitem {Tol} M. Toller , gr-qc 9605052

\bibitem {Rov} C.Rovelli, Class. Quant. Grav. 8, 317,(1991)
 
\bibitem {Unr} W.G.Unruh, R.M.Walde Phys. Rev. D40, 2598, (1989) ,
 H.Kitada Nuov. Cim. 109B , 281 (1995) , gr-qc/9708055

\bibitem {Dew} B. DeWitt , 'Gravitation : an Introduction to
Current Research' ed. L. Witten , (Wiley, N-Y, 1962) 

\bibitem{May2} S.N. Mayburov , in 'Proc.
of  6th Quantum Gravity Seminar', Moscow, 1995 (W.S.,Singapore,1997)
,Proc. of 8th Marsel Grossman Conference', Jerusalem, 1997
(W.S., Singapore, 1998) , eprint gr-qc 9705127 , quant-ph/9801075

\bibitem{May} S.N.Mayburov,  Int. Journ. Theor. Phys. 34,1587 (1995)
,ibid. 37, 401 (1998) 

\bibitem {Desp} D'Espagnat Found. Phys. 20,1157,(1990)

\bibitem {Schw} S. Schweber 'An Introduction to Relativistic Quantum
 Field Theory' ,(New-York,Row-Peterson ,1961) 

\bibitem {Dir} P.A.M.Dirack 'Lectures on Quantum Mechanics' , 
(New-York, Yeshiva University ,1964) 

\bibitem {Git} D.Gitman , I.Tyutin 'Canonical Quantization of Fields with
Constraints', (Berlin, Springer, 1990)
 
\bibitem {Bar} A.Barut, R.Ronczka 'Theory of Group Representations
and Applications', (PWN,Warszawa,1977)

\bibitem {Git2} D.Gitman , I.Tyutin  Class. Quant.  Grav. 7 ,2131, (1990),
 JETP Letters 51,4(1990)

%\bibitem{Blo} D.Blokhintsev 'Microscopic space-time'(Moscow,Nauka,1982)

\bibitem {Per} A.Peres Am. Journ. Phys. 48, 552 (1980)

\bibitem {Hol} A.Holevo 'Probabilistic and Statistical Aspects of 
Quantum Theory', (North-Holland,Amsterdam,1982)

 

\bibitem {Fock} V.Fock Sow. Phys. 12, 404, (1937)


%\bibitem {Hal} L.Khalfin Uzpehi 160,185,(1990)


\bibitem{Wig} E.Wigner, T.Newton Rev. Mod. Phys. 21,400,(1949)

\bibitem {Aha2} Y.Aharonov, D.Albert Phys. Rev. D 24 ,359 (1981)
 
\bibitem{Coe} F.Coester Helv. Phys. Acta 38, 7 (1965)

\bibitem {Byc} E.Byckling, K.Kajantie 'Particles Kinematics'
  (London, John Wiley and Sons ,1973)

\bibitem {Lan} L.Landau, E.Lifshitz 'Classical Field Theory'
(Moscow , Nauka ,1981)

%\bibitem {Bus} P. Busch, M. Grabowski, P.J. Lahti
%Phys. Lett. A 191, 357 (1994),

\bibitem {Byc} E.Byckling, K.Kajantie 'Particles Kinematics'
  (London, John Wiley and Sons ,1973)

\bibitem {Lan}  L.Landau , E.Lifshitz
        'Statistical Physics' , ( Moscow , Nauka ,1979)

%\bibitem {QE} A.Akhiezer , V.Berestetsky ,'Quantum Electrodinamics'
         (Moscow ,Nauka ,1981)

\bibitem {Bro} J.Brown K.Kuchar Phys. Rev. D51 ,5600 (1995);
 ibid D43, 419 (1991)
 




%\bibitem {Zur} W.Zurek Phys.Rev. D26,1862,(1982)
\end{thebibliography}
\end{document}